\documentclass[graybox, natbib]{svmult}
\bibpunct{(}{)}{;}{a}{}{,}

\usepackage{mathptmx}       
\usepackage{helvet}         
\usepackage{courier}        

\usepackage{makeidx}         
\usepackage{graphicx}        

\usepackage{multicol}         
\usepackage[bottom]{footmisc} 

\usepackage[normalem]{ulem}   

\usepackage{url}              

\usepackage{soul}             

\newcommand{\be}{\begin{equation}}
\newcommand{\ee}{\end{equation}}
\newcommand{\tensor}[1]{\stackrel{\leftrightarrow}{#1}}

\begin{document}

\title*{Polymer Solutions}

\author{Burkhard D\"unweg}

\institute{Burkhard D\"unweg
\at
Max Planck Institute for Polymer Research,
Ackermannweg 10,
55128 Mainz, Germany;
Department of Chemical Engineering,
Monash University,
Clayton, VIC, Australia;
\email{duenweg@mpip-mainz.mpg.de}
}

\maketitle

\abstract{The article provides a brief general introduction into the
  concepts of scaling, universality, and crossover scaling, plus the
  blob concept that provides an intuitive picture of crossover
  phenomena. We present the most important static and dynamic scaling
  laws for unentangled uncharged polymer solutions, together with
  their test and refinement by careful computer simulations. A hoard
  of simulation methods has been developed for these systems, and
  these will be briefly discussed as well.}

\section{Introduction}
\label{sec:Intro}

Polymers exist in a variety of states and situations. They may appear
as bulk systems or in confined geometries (like films) and they may be
solid (semi--crystalline, rubbery, or glassy) or liquid. In the liquid
state the macromolecules may form a dense melt, or they are dissolved
in a solvent of good or poor quality. Finally, for a solution it makes
a huge difference if the molecules are charged or not; in the former
case one has a polyelectrolyte system.

Theoretical physics is mainly interested in the \emph{universal}
properties of polymer systems, i.~e. those properties that do not
depend on the details of the chemistry that defines the
mo\-no\-mer. Lots of insight has been gained by deliberately
discarding these details and rather focusing on the effects that
result from the physics of macromolecules as such. The most important
aspects here are (i) connectivity, i.~e. the macromolecular
architecture, which may be linear chains, rings, stars, combs,
networks, etc. (in other words, the topology arising from bonded
interactions); (ii) non--bonded interactions (here in particular the
excluded--volume interaction, solvent quality effects, and possibly
long--range electrostatics); (iii) (possibly) geometric restrictions;
and (iv) (for solutions) the dynamic coupling between the motion of
macromolecules and the flow of the surrounding solvent (the so--called
``hydrodynamic interaction''). This is altogether the huge field of
theoretical polymer physics, for which excellent
textbooks~\citep{gennes_scaling_1979,doi_theory_1988,grosberg_statistical_1994,rubinstein_polymer_2003}
exist. Even though the complicated chemistry has been replaced by
simplified or coarse--grained models, understanding the physics is
still a challenging and complicated problem, which one cannot simply
``solve'' by straightforward pencil--and--paper analytical
theory. Rather, one has to rely on a combination of intuitive insight,
theoretical approximations, experiments in the laboratory, as well as
careful numerical studies of well--defined models.

The most successful computer models in the ``universal'' regime of
phenomena are (i) simple lattice models and (ii) bead--spring models
in the continuum. Both types of models can faithfully model
connectivity and interactions (aspects (i) and (ii) of the previous
paragraph); however, they have different strengths and weaknesses when
it comes to further aspects. Lattice models are particularly
well--suited for Monte Carlo studies of static properties, while
bead--spring models are particularly amenable for studying the
dynamics of systems with hydrodynamic interactions, and for applying
Molecular Dynamics and similar methods, which are much easier to
parallelize than Monte Carlo algorithms.

The present article attempts to provide some overview of the physics
(statics and Brownian dynamics) of \emph{neutral} polymer solutions in
the \emph{bulk}, and computer simulations that have provided
confirmation and/or refinement of the underlying ideas. For
polyelectrolyte solutions and dense melts, please see the
contributions by C. Holm and G. S. Grest in this volume, respectively.

\section{Scaling laws}
\label{sec:Scaling}

A good deal of theoretical polymer physics is concerned with
so--called ``scaling laws''. This is a very general concept, which has
proven extremely useful not only for polymers, but also, e.~g., in the
theory of critical phenomena, or the study of turbulence. From an
abstract point of view, scaling laws are nothing but a special case of
the general observation that the mathematical structure of a physical
theory is strongly restricted or perhaps even fully determined by the
underlying symmetries. Here we deal with a special symmetry, which is
the invariance under the rescaling of parameters. Suppose we consider
a certain physical quantity $Q$ and we are interested in its
dependence on another physical quantity $P$. As an example, let us
think about the dependence of the average size $R$ of a polymer coil
on the contour length $L$ of the (linear) molecule. Let us further
assume that we pick a certain value of $P$, $P = p$, as the basic unit
for $P$. Let $q$ be the value of $Q$ for this particular $P$
value. Then we may write the relation in dimensionless form,
\be
\label{eq:GeneralLaw}
\frac{Q}{q} = F \left( \frac{P}{p} \right) ,
\ee
where $F$ is a dimensionless function with $F(1) = 1$. Of course, we
could also use a different unit system, by picking a different value
$p'$ for the $P$ units, and the corresponding value $q'$ for the $Q$
units, such that we also have
\be
\label{eq:Rescaled1}
\frac{Q}{q'} = \frac{Q}{\lambda q}
= G \left( \frac{P}{p'} \right) ,
\ee
with (in general) another function $G$, $G(1) = 1$, where the
rescaling factor is given by $\lambda = q' / q$. On the right--hand
side we can similarly introduce $\phi = p' / p$.  Obviously $\phi$
will depend on $\lambda$, $\phi = \phi(\lambda)$. Now, the system is
\emph{scale invariant} if $\phi$ depends \emph{only} on $\lambda$ but
not on the point from where the rescaling started (i.~e. not on $p$,
$q$), and if also $G = F$. In other words, scale--invariant systems
are those that are characterized by a \emph{lack} of intrinsic scale,
or those where the physics does \emph{not} provide some most natural
unit system. In our polymer coil example this means that it does not
matter how long the subchains are into which the full molecule is
decomposed (as long as these subchains are long compared to the size
of a chemical mo\-no\-mer, and short compared to the overall contour
length). Since we can combine two rescaling transformations with
factors $\lambda$ and $\mu$ into a single one with factor $\lambda
\mu$, we have, for scale--invariant systems,
\be
\label{eq:FunctionalEquation}
\phi(\lambda \mu) = \phi(\lambda) \phi(\mu)
\ee
and of course $\phi(1) = 1$. By mapping this relation onto a
differential equation, it is trivial to show that its solution is a
power law,
\be
\label{eq:OriginalPowerLaw}
\phi(\lambda) = \lambda^{1/\alpha} ,
\ee
with an undetermined exponent $\alpha$. Insertion into
Eq.~\ref{eq:Rescaled1} yields
\be
\frac{Q}{\lambda q} = F \left( \frac{P}{\lambda^{1/\alpha} p} \right) ,
\ee
or, with $x = P/p$,
\be
\label{eq:ScaleInvarianceOneArgument}
F(x) = \lambda F(\lambda^{-1/\alpha} x) .
\ee
By picking the special value $\lambda = x^\alpha$, we see that $F$ is
also a power law,
\be
F(x) = x^\alpha .
\ee
In other words, scale invariance automatically gives rise to power
laws. For our polymer coil example this means
\be
\frac{R}{R_0} = \left( \frac{L}{L_0} \right)^\nu ,
\ee
where $R_0$ is the coil size for contour length $L = L_0$, and the
Flory exponent $\nu$ depends on the physical situation. For flexible
(i.~e. sufficiently long) polymers in three dimensions $\nu$ takes the
value $1/2$ for random--walk (RW) statistics, while it is roughly
$0.59$ for self--avoiding--walk (SAW) statistics, which applies in
good--solvent conditions. Finally, for a chain that forms a collapsed
globule due to attractive interactions, $\nu = 1/3$.

It is important to realize that the exponents of scaling laws are
typically \emph{universal} (and this is certainly true for
$\nu$). This is so because scale invariance means that the system
``looks the same'' after proper rescaling. Now, the idea of the
renormalization group for
polymers~\citep{cloizeaux_polymers_1991,schafer_excluded_1999} is that
one should start from an original system and then subject it to a
coarse--graining procedure, where several original mo\-no\-mers are
lumped into new effective mo\-no\-mers. Iterating this, the chain more
and more ``forgets'' its chemical details, while only the asymptotic
scale--invariant structure remains --- and this is the same for all
original systems within a so--called ``universality class''.  For
polymers, all chains with relevant excluded--volume interactions
belong to the universality class of SAWs, while those with turned--off
excluded volume to the RW universality class. For simulations, the
concept of universality implies that any model can in principle be
used, as long as it falls into the universality class that one wishes
to study. This in turn means that the construction of models is mainly
guided by considerations of conceptual simplicity, computational
efficiency, and convenience in general.

Further important universal quantities are amplitude ratios and
crossover scaling functions. The latter will be discussed in the next
section; the former are simply the ratios of prefactors of scaling
laws in dimensionless form. For example, one can study various
measures of the size of a polymer coil, i.~e. the end-to-end-distance
$R_E \equiv \left< R_E^2 \right>^{1/2}$, the gyration radius $R_G
\equiv \left< R_G^2 \right>^{1/2}$, and the hydrodynamic radius $R_H
\equiv \left< R_H^{-1}\right>^{-1}$, with
\begin{eqnarray}
R_G^2 & = & N^{-1} \sum_i \left( \vec r_i - \vec R_{CM}\right)^2 ,
\\
\vec R_{CM} & = & N^{-1}  \sum_i \vec r_i ,
\\
\vec R_E & = & \vec r_N - \vec r_1 ,
\\
R_H^{-1} & = & N^{-2} \sum_{i \ne j}
\left\vert \vec r_i - \vec r_j \right\vert^{-1} .
\end{eqnarray}
Here $N$ is the number of mo\-no\-mers of the chain, whose coordinates
are denoted with $\vec r_i$. In the asymptotic long--chain limit the
ratios $R_E / R_G$ and $R_G / R_H$ are universal numbers, taking the
values $\sqrt{6}$ and $8 / (3 \sqrt{\pi})$ for three--dimensional RWs.

\section{Crossover scaling}
\label{sec:CrossoverScaling}

There are many situations where one needs to consider the dependence
of a quantity on more than a single variable. For example, in polymer
solutions one is interested in the dependence of the coil size on
degree of polymerization $N$, concentration $c$ (total number of
mo\-no\-mers per unit volume), and solvent quality. In such a
situation, scale invariance is expressed by a straightforward
generalization of Eq.~\ref{eq:ScaleInvarianceOneArgument}:
\be
F(x_1, x_2, \ldots) = \lambda F(\lambda^{-1/{\alpha_1}} x_1,
\lambda^{-1/{\alpha_2}} x_2, \ldots) .
\ee
A particularly important case occurs if there are just two arguments,
in which case we have
\be
F(x_1, x_2) = x_1^{\alpha_1}
F(1, x_1^{-{\alpha_1}/{\alpha_2}} x_2) ,
\ee
such that apart from the power law $x_1^{\alpha_1}$ we also have a
dependence on the ``crossover scaling variable'' $x_c \equiv
x_1^{-{\alpha_1}/{\alpha_2}} x_2$, while $F(1,
x_1^{-{\alpha_1}/{\alpha_2}} x_2)$ is then called a ``crossover
scaling function''. Typically the behavior becomes simple in the
asymptotic limits $x_c \gg 1$, $x_c \ll 1$, where simple power laws $F
\propto x_1^{\beta_>}$, $F \propto x_1^{\beta_<}$ are recovered. In
such a case, the crossover scaling function must itself asymptotically
behave like an appropriate power law. Finally, if $F$ describes the
behavior of a universal ratio, it must itself be universal in the
limit of long chains.

\section{Blobs}
\label{sec:Blobs}

Crossover phenomena in polymer solutions can conveniently be described
in terms of so--called ``blobs''. A blob is a portion of the polymer
chain that is composed of $g$ mo\-no\-mers and has a typical extension
(the ``blob size'') $\xi$. This length scale marks the crossover
between two different behaviors, and typically an energy of $k_B T$
(thermal energy) is associated with it. The blob concept provides a
nice pictorial description of crossover phenomena and is hence a very
useful tool for deriving crossover scaling laws. The most important
crossovers in the statics of polymer solutions are those between RW
and SAW behavior, driven by (i) attractive effective interactions and
(ii) concentration.

For a single isolated chain, the quality of solvent can be measured in
terms of an effective interaction energy $\epsilon (T)$, which
measures the temperature--dependent degree of attraction between two
mo\-no\-mers. At the temperature of the Theta
transition~\citep{lifshitz_problems_1978,schafer_excluded_1999} ($T =
\Theta$) the repulsive and attractive parts of the interaction cancel
out, such that effectively the chain behaves as a RW. In the vicinity
of $T = \Theta$ we may write $\epsilon (T) = \epsilon_0 (1 - \Theta /
T)$, which gives rise to a dimensionless interaction parameter
$z^\star = (k_B T)^{-1} \epsilon_0 (1 - \Theta / T)$. For any $z^\star
> 0$ the chain structure is asymptotically a SAW. However, if
$z^{\star}$ is small, the amount of repulsion is too small to disturb
the RW statistics on small length scales. This gives rise to a thermal
blob size $\xi_T$ corresponding to $g_T$ mo\-no\-mers, of which each
has a size $b$, such that $\xi_T \sim b g_T^{1/2}$. The number of
mo\-no\-mer--mo\-no\-mer contacts within the blob is estimated as
$g_T^{1/2}$. The blob size is found by equating the total energy in
the blob with $k_B T$, i.~e. $g_T^{1/2} z^\star \sim 1$, or $\xi_T
\sim b / z^\star$. Visualizing a very long chain as a SAW composed of
RW blobs, one finds $R \sim \xi_T (N / g_T)^\nu$ or
\be
\label{eq:RasymptoticTheta}
R \sim b N^{1/2} \left( N^{1/2} z^\star \right)^{2 \nu - 1} ,
\ee
from which the relevant crossover scaling variable $z = N^{1/2}
z^\star$ is read off. For chains that violate the condition $N \gg
g_T$, Eq.~\ref{eq:RasymptoticTheta} is generalized to
\be
R \sim b N^{1/2} f(z) ,
\ee
where the crossover scaling function $f(z)$ behaves like $f(z) \sim
z^{2 \nu - 1}$ for $z \gg 1$, while $f(z) \sim 1$ for $z \ll 1$.

\begin{figure}[t]
\includegraphics[scale=.45]{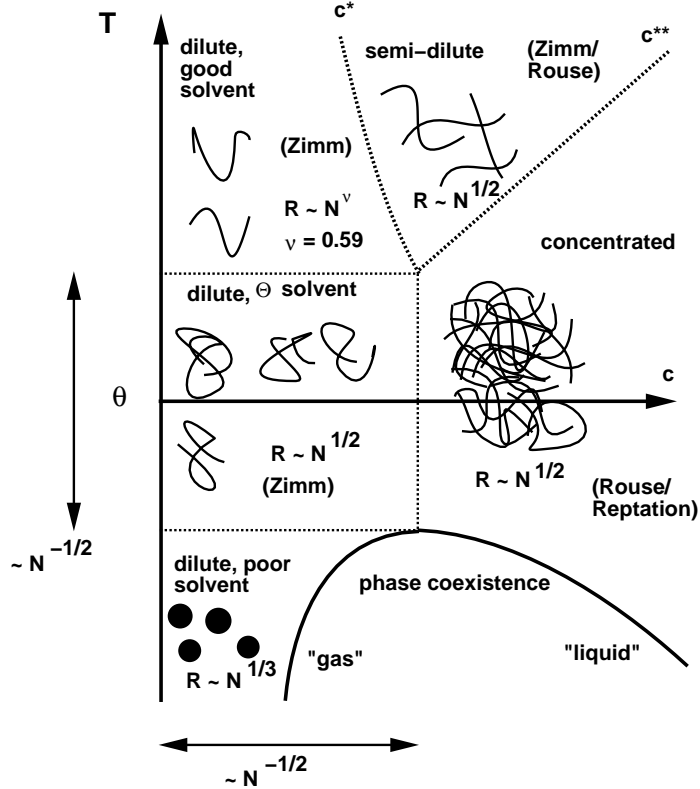}
\caption{Generic phase diagram of a polymer solution}
\label{fig:PhaseDiagram}
\end{figure}

Next, let us consider the concentration--driven crossover from SAW to
RW behavior in perfectly good solvent, as a result of Flory screening
of excluded--volume
interactions~\citep{gennes_scaling_1979,rubinstein_polymer_2003}. The
overlap concentration $c^\star$ is the concentration where an
arrangement of perfectly swollen chains is just space--filling,
i.~e. $c^\star \sim N / R^3 \sim b^{-3} N^{- (3 \nu - 1)}$. At
concentrations $c$ that significantly exceed $c^\star$, the chains
overlap. This gives rise to concentration blobs of size $\xi_c$,
containing $g_c$ mo\-no\-mers. Since on length scales below $\xi_c$
there is no overlap, the SAW structure is unperturbed in this
regime. Conversely, on scales above $\xi_c$ we have RW
behavior. Therefore, $\xi_c \sim b g_c^\nu$ and $c \sim b^{-3} g_c^{-
  (3 \nu - 1)}$, or $\xi_c \sim b (b^3 c)^{- \nu / (3 \nu - 1)}$. A
long chain is then a RW composed of SAW blobs, $R \sim \xi_c
(N/g_c)^{1/2}$, or
\be
\label{eq:ConcentrationCrossoverR}
R \sim b N^\nu \left( \frac{c}{c^\star} 
\right)^{- \frac{\nu - 1/2}{3 \nu - 1} } ,
\ee
from which the natural crossover scaling variable $c / c^\star$ is
read off. The generalization in terms of a crossover scaling function
is $R \sim b N^\nu f(c / c^\star)$, where now $f(c / c^\star) \sim 1$
for $c / c^\star \ll 1$, while for $c / c^\star \gg 1$ the power law
Eq.~\ref{eq:ConcentrationCrossoverR} is recovered. A solution whose
mo\-no\-mer concentration is small but whose chains are so long that
there is nevertheless a strong overlap is called ``semidilute''. The
semidilute regime ends at a concentration $c^{\star \star}$ where the
blob size has shrunken to the mo\-no\-mer size, such that no SAW
regime is left.

In the general case, the structure is determined by the interplay
between concentration and solvent quality effects, or the competition
between $\xi_T$ and $\xi_c$. For the ideal picture of a semidilute
solution, we have $b \ll \xi_T \ll \xi_c \ll R$. On length scales
below $\xi_T$ and above $\xi_c$ we have RW behavior. This is due to
attraction on the small scales and due to Flory screening on the large
scales. For length scales between $\xi_T$ and $\xi_c$ we have SAW
behavior; this regime shrinks more and more upon deteriorating the
solvent quality or upon increasing the concentration. This picture
gives rise to the generic phase diagram shown in
Fig.~\ref{fig:PhaseDiagram}. For more details on the derivation, see
the textbook literature, or the Supplemental Material of
\citep{jain_dynamic_2012}. It is important to note that all universal
ratios can, in the asymptotic limit of infinite chain length, be
expressed in terms of just the two crossover scaling variables $z$ and
$c/c^\star$.

\section{Dynamic scaling}
\label{sec:DynamicScaling}

Polymer statics provides us with two important length scales, the
mo\-no\-mer size $b$ and the coil size $R$. In case the system needs
to be described in terms of a blob picture, there may be a blob size,
or possibly even more blob sizes, as additional important length
scales. In a dense melt of long chains, there may also be
entanglements, which give rise to a ``tube diameter'' as yet another
important length scale. In the present article, we will only consider
non--entangled systems, where the chains are either too short or too
dilute to develop entanglements.

The idea of dynamic scaling for the Brownian motion of polymers may
then be understood as follows: For each length $l$, there is an
associated time $\tau(l)$. This time may be viewed as the time that a
sub--chain of extension $l$ (in real space, \emph{not} along the
contour) needs to move diffusively by its own size. Alternatively, we
may also pick a single mo\-no\-mer and study the time dependence of
its root mean square displacement, $\left< (\Delta \vec r)^2
\right>^{1/2}$. The time $\tau(l)$ would then be given by the time that
passes until $\left< (\Delta \vec r)^2 \right>^{1/2} = l$. The
underlying concept is here that the single--mo\-no\-mer motion must be
consistent with the motion of the object as a whole. Now suppose we
consider $l$ values that are well within a regime that is bounded by
two of the important scales mentioned in the previous paragraph, with
no further important scale in between. Then we again have a lack of a
naturally provided unit system, and this applies not only to lengths
but also to times. Therefore, the dynamics in such a regime must be
described by a power law
\be
\label{eq:DynScaling}
\tau(l) \propto l^z ,
\ee
which defines the \emph{dynamic exponent} $z$ (not to be confused with
the crossover scaling variable of the previous chapter!), applicable
to the regime under consideration.

Let us first consider a system where no blobs occur. This can either
be a dilute solution ($c \ll c^\star$) or a dense melt ($c \gg
c^{\star \star}$). The longest relaxation time $\tau_R$ is associated
with the coil size,
\be
\tau_R \propto R^z .
\ee
Dynamic scaling then implies that the mean square displacement for
times $t$ with $\tau(b) \ll t \ll \tau_R$ must obey
\be
\left< (\Delta \vec r)^2 \right> \propto t^{2/z} .
\ee
Furthermore, if $D$ denotes the center--of--mass diffusion constant of
the chain as a whole, we have $D \tau_R \sim R^2$ or $D \propto
R^{-(z-2)}$.

For an isolated chain in solvent (dilute limit), the intra--chain
hydrodynamic interaction (i.~e. the strong coupling of the mo\-no\-mer
motion to the flow of the surrounding solvent) gives rise to
so--called Zimm dynamics~\citep{doi_theory_1988}. A characteristic
feature of Zimm dynamics is that the diffusive behavior of the chain
is essentially that of a Stokes sphere, $D \sim (k_B T) / (\eta R)$,
where $\eta$ is the solvent viscosity. More precisely, the approximate
Kirkwood theory~\citep{doi_theory_1988} predicts $D = (k_B T) / (6 \pi
\eta R_H)$. From this, one reads off $z = 3$.

In case the chain is in a dense melt, the hydrodynamic interactions
are screened (more about this below). Therefore there are no
hydrodynamic correlations in the mo\-no\-mer displacements, and if
entanglements play no role (which is the case if the chains are not
too long), then the relevant theory for the dynamics is the Rouse
model~\citep{doi_theory_1988}. Here one simply assumes a homogeneous
viscous background and each mo\-no\-mer has a friction constant
$\zeta$. The friction coefficients of these mo\-no\-mers simply add up
to yield the total friction coefficient of the chain. Therefore $D =
(k_B T) /(N \zeta)$. Since $R \sim b N^{1/2}$ (in a dense melt we have
RW statistics), $D \propto R^{-2}$ or $z = 4$.

In a semidilute solution, the length scale that governs the crossover
from Zimm to Rouse dynamics is again the blob size $\xi_c$. Up to this
length scale, one has unscreened excluded--volume and hydrodynamic
interactions, i.~e. SAW statistics and Zimm dynamics. The
corresponding time scale is the blob relaxation time $\tau_\xi \sim
\eta \xi_c^3 / (k_B T)$. The blob as a whole has a Stokes friction
coefficient $\sim \eta \xi_c$. Computer
simulations~\citep{ahlrichs_screening_2001} have shown that
hydrodynamic interactions are unscreened as long as the \emph{time}
scale is significantly below $\tau_\xi$, regardless of length
scales. This is reasonable, since on these short time scales all
mo\-no\-mers just move with the flow, such that correlations exist
even on length scales significantly beyond $\xi_c$. Conversely, for
times significantly above $\tau_\xi$, the blobs ``feel'' that they are
not free to move (or that chain--chain collisions occur). Therefore,
the blobs then dampen the hydrodynamic flow velocity $\vec u$ with a
friction force per unit volume of~\citep{de_gennes_dynamics_1976}
$\sim \eta \xi_c \xi_c^{-3} \vec u$, which is a term that should be
added to the Stokes equation for $\vec u$. For such a situation, the
flow field generated by a point force does no longer decay like $1/r$
($r$: distance from the point force) but rather like
$\exp(-r/\xi_H)/r$, where the hydrodynamic screening length $\xi_H
\sim \xi_c$. Hydrodynamic screening is thus understood as a
randomization of hydrodynamic correlations, induced in essence by
chain--chain collisions.

It should also be mentioned that melts do not exhibit strict Rouse
dynamics in the dense limit, even if the chains are quite short. The
reason is dynamic coupling of the chain motion to the viscoelastic
modes of the ``matrix'', which gives rise to subtle
corrections~\citep{farago_mode-coupling_2012a,farago_mode-coupling_2012b}.
Similarly, there are also subtle corrections to the RW statistics of
polymer chains in a melt~\citep{wittmer_long_2004}. Both results have
been obtained by careful computer simulations.

\section{Simulations I: Statics}
\label{sec:SimulationsI}

To study universal static single--chain properties, the method of
choice is clearly Monte Carlo (MC) of lattice models, where chains are
simply walks on a lattice. Mostly simple--cubic lattices are studied,
but other lattice structures are permitted as well. The so--called
pivot algorithm~\citep{madras_pivot_1988} is presently the most
efficient method known. Here one randomly selects a sub--chain and
rotates it by a random angle around a random axis. This is an MC trial
move, which is accepted or rejected by the standard Metropolis
criterion. Presently the fastest--known implementation is that by
N. Clisby~\citep{clisby_accurate_2010}, where the information about
the chain conformation is stored in a somewhat unconventional manner:
Firstly, one defines a bounding box about the chain as a whole.
Associated with it are global properties like number of mo\-no\-mers,
end--to--end--vector, gyration radius, center--of--mass coordinate.
Then one subdivides the chain into two sub--chains, for each of which
the analogous information is stored. This is done recursively all the
way to the mo\-no\-mer level, such that one obtains a binary tree. The
coordinates of each box and its contents are stored \emph{relative} to
the coordinates of the coarser level. This makes it possible to move
one such ``container'' as a whole without the need to ever touch the
data of the finer levels that it contains. Similarly, overlap checks
are done by checking the overlap between bounding boxes: If they do
not overlap, then their contents will surely not overlap either. With
such tricks it is possible to reduce the computational complexity of
one pivot move to $\ln N$, such that very long chains are
accessible. A recent study~\citep{clisby_high-precision_2016} has thus
been able to find for three--dimensional SAWs: $\nu = 0.58759700(40)$,
$R_G / R_H = 1.5803940(45)$, $R_E^2 / R_G^2 = 6.253531(10)$.

Similarly, accurate MC calculations have been able to study the
$\Theta$ transition in three
dimensions~\citep{grassberger_pruned-enriched_1997} and in particular
investigate the subtle logrithmic corrections to scaling that occur
there. For this study the so--called ``PERM'' (prune-enriched
Rosenbluth method) algorithm was used, where chains are grown step by
step, and statistical criteria decide at each step if a chain is
terminated, continued, or even branched to generate yet another
chain. In this way, unbiased samples of long chains may be generated.
Universal crossover scaling functions for the $\Theta$ transition were
studied as well~\citep{kumar_equilibrium_2003}, using the methodology
of Brownian Dynamics (BD; we will discuss this method briefly
below). This study emphasized the importance of appropriate
extrapolation procedures: In order to find the crossover scaling
function, one should work at a constant value of the crossover scaling
variable (here $z = N^{1/2} z^{\star}$) and study the residual
dependence of a universal ratio (like $R_G(T) / R_G(T = \Theta)$) on
the chain length. This residual dependence is a correction to scaling;
therefore the asymptotic universal behavior is obtained after
extrapolation $N \to \infty$.

The crossover scaling for Flory screening in good solvent was studied
utilizing a lattice model and MC
simulation~\citep{paul_crossover_1991}. However, here the model was
the so--called bond fluctuation model~\citep{carmesin_bond_1988},
where mo\-no\-mers do not occupy single sites but rather elementary
cubes, while the connecting bonds may vary within limits. This allows
to implement a MC dynamics that involves simply a random displacement
of an elementary cube on the lattice. Indeed it was found that a
crossover from SAW to RW statistics occurs, with a crossover length
scale $\xi_c$ that exhibits the concentration dependence predicted by
blob theory. Later, the same model was also used to reveal the
corrections to purely Gaussian behavior in a
melt~\citep{wittmer_long_2004}.

Finally, the double crossover that results from the competition
between the Theta blobs and the Flory screening blobs has recently
been studied by BD~\citep{jain_dynamic_2012}. The internal blob
structure could not be resolved, for lack of sufficiently long chains;
however the dependence of total--chain properties like the coil size
on concentration and solvent quality was in perfect agreement with
blob theory. The same was true for dynamic properties like the
diffusion constant.

\section{Simulations II: Dynamics}
\label{sec:SimulationsII}

For studies of dynamics, one needs algorithms that faithfully
reproduce the motion of the mo\-no\-mers, at least on the (typically
long) time scales that one is interested in. Obviously, brute--force
Molecular Dynamics (MD) will satisfy this condition, if it involves
all particles in the system. This approach has been highly successful
for the studies of melts (see contribution by G.~S. Grest in this
volume), and can in principle also be applied to the dynamics of
solutions, where the hydrodynamic interactions are taken into account
by explicit solvent particles. Zimm dynamics for a single chain could
thus be successfully
established~\citep{pierleoni_molecular_1992,dunweg_molecular_1993}.
For such studies of a single macromolecule in a simulation box one has
to take into account that the latter is typically not much larger than
the chain itself. Therefore, one has to deal with strong finite--size
effects, which scale as $R_G / L$, where $L$ is the linear box
size. This is a direct consequence of the long--range nature of the
hydrodynamic interactions: Since the flow field generated by a point
force decays like $1/r$ ($r$: distance from the force center), the
correlations between the stochastic displacements of two distinct
mo\-no\-mers are proportional to the inverse interparticle
distance. The theory thus provides detailed quantitative predictions
about the magnitude of such finite--size effects, and this in turn
makes it possible to quantitatively check Zimm theory even in a
finite--box situation.

However, for solutions MD is nearly always unnecessarily expensive and
can rather be replaced by cheaper algorithms that simulate the solvent
degrees of freedom in a simplified fashion. The only situations where
this is not true are either fairly concentrated solutions, where the
solvent contribution to the computational effort is only moderate, or
investigations of local atomistic dynamics, where local packing and
similar phenomena are of specific interest. In all other cases, the
effect of the solvent can be summarized by (i) its quality, which may
be modeled by just a suitable effective momomer--mo\-no\-mer
interaction, and (ii) the hydrodynamic interactions, which give rise
to dynamic correlations between mo\-no\-mers as a result of momentum
transport through the solvent. The crucial observation is here that
the solute--solvent system is characterized by a large separation of
time scales: The slowest degree of freedom in the solvent is diffusive
momentum transport, characterized by the ``kinematic viscosity''
$\eta_{kin} = \eta / \rho$, i.~e. the ratio between shear viscosity
and mass density, which has the dimension of a diffusion constant. The
dimensionless ``Schmidt number'' $Sc = \eta_{kin} / D$ then relates
this to the diffusion constant $D$ of an immersed particle, or, more
generally, to the diffusion constant of some immersed soft--matter
object of size $R$. Typically, in dense fluids $Sc \gg 1$ even for
solvent particles, due to a sizeable viscosity value --- note
$\eta_{kin} \propto \eta$ but $D \propto 1 / \eta$. For large
(isolated) macromolecules, the corresponding Schmidt number is yet
much larger, as $Sc \propto 1 / D \propto R$. For these reasons, we
may \emph{either} replace the solvent by some sort of ``generalized
hydrodynamics solver'', i.~e. a set of more or less artificial degrees
of freedom that exhibit the correct hydrodynamic behavior on large
length and time scales, \emph{or} dispose of the solvent altogether,
by assuming that the flow field follows the configuration of
mo\-no\-mers instantaneously, such that it becomes completely enslaved
to the latter, and thus no longer appears as an explicit degree of
feedom.

Let us begin with the latter approach. This is the realm of
\emph{Brownian Dynamics}~\citep{ottinger_stochastic_1995}. Here one
solves a discretized stochastic differential equation for the
mo\-no\-mer coordinates $\vec r_i$, using a finite time step $h$. The
update rule can then be written as
\be
  \label{eq:BDalg}
  r _{i \alpha} (t + h) = 
  r_{i \alpha} (t)
  + h \sum_j \mu_{i \alpha, j \beta} F_{j \beta}
  + k_B T h \sum_j \partial \mu_{i \alpha, j \beta} / \partial r_{j \beta}
  + \sqrt{2 k_B T h} \sum_j \sigma_{i \alpha, j \beta} q_{j \beta} .
\ee
Here Greek letters indicate Cartesian indexes with Einstein summation
convention. $\vec F_j$ is the force acting on particle $j$, while
$\tensor \mu_{ij}$ is the mobility tensor that describes the
hydrodynamic correlations between the mo\-no\-mers $i$ and $j$.
Typically, the Rotne--Prager tensor~\citep{ottinger_stochastic_1995}
is used. In case one is not interested in correct solution dynamics,
one may simply turn the hydrodynamic interactions off and replace
$\tensor \mu_{ij}$ with a multiple of the unit tensor. In this case,
the method will produce Rouse--like dynamics for a single--chain
simulation. The last term of Eq.~\ref{eq:BDalg} denotes the stochastic
displacements, where $q_{i \alpha}$ are random variables with
\begin{eqnarray}
\left< q_{i \alpha} \right> & = & 0 \\
\left< q_{i \alpha} q_{j \beta} \right> & = &
\delta_{ij} \delta_{\alpha \beta} ,
\end{eqnarray}
while the matrix $\sigma_{i \alpha, j \beta}$ satisfies
\be
\sum_{k} \sigma_{i \alpha, k \gamma}
                \sigma_{j \beta, k \gamma} =
\mu_{i \alpha, j \beta} .
\ee
This approach was pioneered by a seminal paper nearly forty years
ago~\citep{ermak_brownian_1978} and has seen many refinements since
then. The main difficulties are the evaluation of the mobility tensor,
which couples all mo\-no\-mers in the system, and the calculation of a
suitable square root. These problems have remained a computational
challenge for decades. Standard Ewald sums for multi--chain
systems~\citep{jain_optimization_2012} have met moderate success, but
only recently has a method been published~\citep{fiore_rapid_2017}
whose computational effort scales strictly \emph{linearly} with the
number of involved mo\-no\-mers.

The generalized hydrodynamics solvers are technically much easier and
also much more easy to parallelize. They also scale linearly with the
number of mo\-no\-mers, however at the expense of an additional large
set of explicit solvent degrees of freedom. These solvers all include
thermal fluctuations in some way or another. This is necessary because
in soft--matter physics we are dealing with length and time scales
that are so small that fluctuations play a role. Obviously, Brownian
motion of polymer chains could not be studied if fluctuations were
absent. Therefore such methods are not fully macroscopic but are
rather frequently called ``mesoscale'' methods.

One can distinguish two classes of mesoscale methods, depending on the
way how thermal fluctuations are treated. The first class, which one
may call ``MD--like'', are particle methods where the amount of
thermal fluctuations per degree of freedom is similar to what one
would get in an MD simulation. Peculiar to these methods is the
impossibility to adjust the degree of thermal fluctuations
\emph{independently} of the macroscopic fluid properties that are
relevant for hydrodynamics. Conversely, in the second class, which one
may call ``hydrodynamics--like'', the degree of thermal fluctuations
\emph{can} be adjusted independently of the macroscopic properties.
The degree of thermal fluctuations is here a reflection of the degree
of coarse--graining: The more atomistic particles are lumped into one
mesoscale degree of freedom, the smaller is the amount of thermal
fluctuations per mesoscale degree of freedom --- simply as a result of
Gaussian statistics and the law of large numbers.

As this aspect is typically under--emphasized in the literature, let
us illustrate this by a very simple example, a one--dimensional ideal
gas, which we simulate by MD, augmented by a Lowe--Andersen
thermostat~\citep{lowe_alternative_1999} to bring the system to
thermal equilibrium. This thermostat simply picks, from time to time,
a pair of nearby particles at random. The center--of--mass velocity of
that pair then remains unchanged, while the relative velocity is
chosen at random, using the appropriate equilibrium Maxwell--Boltzmann
distribution, such that the total momentum is conserved. The thermal
(root mean square) velocity of a particle is then $(k_B T / m)^{1/2}$,
where $m$ is its mass. This has macroscopic relevance, since this is
also the speed of sound. Now let us assume that we lump $M$ adjacent
particles into a new mesoscale particle. The new system is then again
an ideal gas, which we wish to simulate with the same method. We then
have two choices concerning the question of the mass of the mesoscale
particles: Either we can assign the value $M m$, which is naively the
correct choice, since the bigger particle should indeed exhibit more
inertia. Moreover, the thermal velocity (i.~e. the amount of thermal
fluctuations) is indeed correctly reduced by the factor $M^{-1/2}$.
However, this comes at the price of also reducing the speed of sound
by the same factor --- and this is a value that we would prefer to
keep constant, in order to maintain the time--scale separation between
immersed soft--matter objects and the sound waves. Therefore one
typically chooses the value $m$, thus keeping the macroscopic
properties intact, but overestimating the degree of thermal
fluctuations. In other words, MD--like methods are typically too
restrictive to permit a fully consistent coarse--graining. This
dilemma is solved by the hydrodynamics--like methods, where thermal
fluctuations are an add--on with adjustable strength to a method that
would also work in the strict macroscopic limit with no fluctuations
whatsoever.

Dissipative Particle Dynamics (DPD)~\citep{espanol_statistical_1995}
is directly derived from MD, which is just augmented by a
momentum--conserving Langevin thermostat. Similarly to the
Lowe--Andersen method, DPD is based upon pairs of nearby particles,
which are however not chosen at random but rather considered in their
totality, at every single time step. The projection of the relative
velocity onto the interparticle axis is dampened by a Langevin
friction. This is compensated by stochastic Langevin forces on the two
partcles that also act along the interparticle axis and add up to
zero. The total momentum is conserved, and the
fluctuation--dissipation theorem (FDT) is satisfied. A generalized
version also thermalizes the velocity components perpendicular to the
axis~\citep{junghans_transport_2008}, however, it is presently not yet
fully understood what effects the implied violation of
angular--momentum conservation has on the hydrodynamics.

Quite often, DPD simulations are run with particles that have fairly
soft interaction potentials. This is done in the spirit of
coarse--graining, which in general leads to such softening of
interactions. It also has a practical implication, since softer
potentials also allow to use a larger time step. The most radical
implementation of that idea is to simply run DPD of an ideal gas as a
solvent for soft--matter objects~\citep{smiatek_tunable-slip_2008}.
Using an ideal gas has a huge advantage: The solvent degrees of
freedom are reduced to their prime function, which is to transmit
momentum through the system, and the equlibrium structure of the
immersed objects is unaltered compared to immersion in vacuum. The
viscosity can nevertheless be adjusted to reflect dense--fluid
conditions, by choosing a sufficiently strong friction.

A yet simpler variant is Multi--Particle Collision Dynamics
(MPCD)~\citep{gompper_multi-particle_2009}. Here the ideal--gas
particles are sorted into cubic cells. In each cell the algorithm
determines its local center--of--mass velocity, and the relative
velocities of the particles with respect to it. The latter are then
subjected to a random rotation. This ``collision step'', which
conserves both the momentum and the kinetic energy, serves to
thermalize the ideal gas and is followed by a standard MD ``streaming
step''.

Both DPD and MPCD are ``MD--like'', with a coupling of the
mo\-no\-mers to the solvent that arises naturally from the setup of
the respective algorithms. We will now turn to the
``hydrodynamics--like'' methods.

Smoothed Dissipative Particle Dynamics
(SDPD)~\citep{espanol_smoothed_2003} has been developed to cure the
abovementioned deficiencies of DPD. The name suggests a closer
proximity to DPD than the method actually exhibits. While DPD comes in
spirit fron MD, as essentially a bottom--up approach, SDPD rather is a
top--down method: Here the starting point is Smoothed Particle
Hydrodynamics (SPH)~\citep{monaghan_smoothed_2005}, which is nothing
but a discretization of the Navier--Stokes equations in terms of
particles. This looks deceptively similar to MD but is fundamentally
different: Firstly, in MD both the equation of state and also the
transport coefficients like the viscosity are an output of the
atomistic model, and must be determined by simulation. Conversely, in
SPH they are input parameters. Secondly, MD particles have as only
properties their coordinates and momenta (and possibly their
orientations and angular momenta). SPH particles, on the other hand,
have additional properties ``on board'' that one could not even
\emph{define} for MD particles beause their nature is genuinely
\emph{thermodynamic} --- volume and entropy, which both change in the
course of time as a result of the dynamics. SDPD adds Langevin noise
to the SPH equations of motion such that the FDT is satisfied.
Although the SDPD particles are thermodynamic objects, it is
nevertheless possible to simply connect a set of them via springs and
thus obtain an immersed polymer chain with the correct large--scale
properties~\citep{litvinov_smoothed_2008}. The polymer--solvent
coupling is therefore as straightforward as for DPD and MPCD.

Instead of discretizing the Navier--Stokes equations in terms of
particles, one may also discretize them via a lattice. One therefore
arrives at standard finite--difference or finite--volume
schemes~\citep{donev_accuracy_2010,balboausabiaga_staggered_2012}.
Again, one may add thermal fluctuations to the equations to satisfy
the FDT.

Finally, one may also simulate hydrodynamics via the Lattice Boltzmann
(LB) \citep{dunweg_lattice_2009} method. Here one solves a linearized
and fully discretized version of the Boltzmann equation known from the
kinetic theory of gases. Space and time are discretized in terms of a
lattice spacing $a$ and time step $h$, respectively. Velocity space is
also discretized and reduced to a small discrete set of velocities
$\vec c_i$. Each lattice site contains a set of real--valued positive
variables $n_i$, which are interpreted as the mass density
corresponding to velocity $\vec c_i$. The mass density $\rho$ and the
momentum density $\vec j$ are then obtained as zeroth and first
velocity moment of the populations,
\begin{eqnarray}
  \rho   & = & \sum_i n_i , \\
  \vec j & = & \sum_i n_i \vec c_i .
\end{eqnarray}
The procedure then begins with a collision step, i.~e. a
re--arrangement of the populations on the site such that mass and
momentum are conserved:
\be
  n_i \to n_i^{\star} = n_i + \Delta_i ,
\ee
where the collision ``operator'' $\Delta_i$ satisfies
\begin{eqnarray}
  \sum_i \Delta_i          & = & 0 ,\\
  \sum_i \Delta_i \vec c_i & = & 0 .
\end{eqnarray}
This is followed by a streaming to the adjacent lattice sites, such
that the total procedure can be written in terms of the Lattice
Boltzmann equation (LBE):
\be
  n_i(\vec r + \vec c_i h, t + h) = n_i(\vec r, t)
  + \Delta_i(\vec r, t) .
\ee
This implies that the discrete velocities must be chosen commensurate
with the lattice. For example, the popular D3Q19
model~\citep{dunweg_lattice_2009} which lives on the
three--dimensional simple--cubic lattice, involves nineteen
velocities, which correspond to the six nearest and twelve
next--nearest neighbors, plus the zero velocity.

The method involves lots of adjustable parameters, like the set of
velocity shells, associated weight coefficients, and various details
of the collision operator. All of these are tuned in order to obtain
the correct Navier--Stokes behavior in the continuum limit, which is
found from the algorithm by subjecting the LBE to an asymptotic
(Chapman--Enskog) analysis. The LBE can therefore be used as a
Navier--Stokes equation solver in its own right. Thermal fluctuations
are introduced by adding a suitably chosen stochastic collision
operator to $\Delta_i$. For further details,
see~\cite{dunweg_lattice_2009}.

It should be emphasized that in all of the abovementioned
``hydrodynamics--like'' methods it is very important to make sure that
the FDT is not only satisfied in the asymptotic continuum limit, but
rather for the algorithm as such. Substantial effort has gone into the
development of methods that do satisfy this condition.

In contrast to particle methods, hybrid methods that involve MD for
the polymer chains and a lattice algorithm for the solvent need
special care for the fluid--particle coupling. A particularly simple
approach is a frictional coupling~\citep{dunweg_lattice_2009}, where
each mo\-no\-mer is assigned a Stokes friction coefficient. Therefore
each mo\-no\-mer is not only subject to the conservative forces coming
from other mo\-no\-mers (and possibly yet other sources) but also to a
friction force and a Langevin stochastic force. The former dampens the
\emph{relative} velocity of the particle with respect to the local
flow field, which is obtained via interpolation from adjacent lattice
sites. The latter is just standard Langevin noise that is needed to
satisfy the FDT. Back--coupling is obtained by interpolating the
thus--resulting momentum transfer back to the lattice and enforcing
momentum conservation. Another possibility is to enforce a stick
boundary condition, either on the surface of an extended
particle~\ \citep{dunweg_lattice_2009} or based upon a point--particle
picture~\citep{usabiaga_inertial_2013}.

At the end of this section, we briefly wish to mention a few studies
that have focused on polymer solution dynamics. Zimm dynamics of a
single chain has been studied by BD by many authors,
e.~g. \cite{fixman_implicit_1986,liu_translational_2003,sunthar_dynamic_2006},
where the last study also investigated the solvent--quality driven
crossover behavior. Single--chain Zimm dynamics was also studied by
LB/MD~\citep{ahlrichs_simulation_1999},
MPCD~\citep{mussawisade_dynamics_2005}, and
SDPD~\citep{litvinov_smoothed_2008}. Not surprisingly, all these
studies yield essentially the same results, and it is even possible to
quantitatively map them onto each other --- this has explicitly been
done for LB/MD vs. pure MD~\citep{ahlrichs_simulation_1999} as well as
for LB/MD vs. BD~\citep{pham_implicit_2009}.

Detailed studies of the concentration--driven crossover from Zimm to
Rouse dynamics have been done by both
LB/MD~\citep{ahlrichs_screening_2001} and
MPCD~\citep{huang_semidilute_2010}. Both confirmed the picture of
hydrodynamic screening as outlined in Sec.~\ref{sec:DynamicScaling},
and the latter paper went even beyond to also study non--equilibrium
behavior.

\section{Summary}

Polymer solution statics and dynamics is a beautiful piece of physics
where progress has been made by analytical theory (in particular
scaling considerations), experiments, and computer
simulations. Improved physical and mathematical insight led to the
development of computer simulation methods that went from simple and
fairly brute--force to more and more sophisticated and
problem--oriented, focussing on the essence of the phenomena one
wishes to study. The author hopes that the present contribution has
given the reader a glimpse on how fruitfully theory and simulations
have worked together in this field. For reasons of both space and also
expertise of the author, the present article has only focused on the
most basic equilibrium phenomena and completely left out the highly
important field of non--equilibrium physics, i.~e. nonlinear polymer
solution rheology, which would be worth yet another article in this
series.

\end{document}